\documentclass{epl2}

\usepackage{epsfig}
\usepackage{graphicx}
\usepackage{dcolumn}
\usepackage{bm}

\title{Role of vacancies and impurities in the ferromagnetism of
semiconducting CaB$_6$}

\shorttitle{Ferromagnetism in CaB$_6$}

\author{Kalobaran Maiti,\footnote{Electronic address:
kbmaiti@tifr.res.in}}

\institute{Department of Condensed Matter Physics and Materials
Science, Tata Institute of Fundamental Research, Homi Bhabha Road,
Colaba, Mumbai - 400005, INDIA}


\abstract{We investigate the influence of impurities and vacancies
in the formation of magnetic moment in CaB$_6$ using full potential
{\it ab initio} band structure calculations. CaB$_6$ is found to be
a band insulator with a band gap of about 0.2 eV. The calculated
results indicate that carbon and oxygen substitution in boron
sublattice, usually expected as impurity in low purity borons, do
not play significant role in the local moment formation. Boron
vacancy in the boron sublattice, on the other hand, leads to the
formation of an impurity band in the vicinity of the Fermi level,
which exhibit finite exchange splitting and magnetic moment. All
these results provide remarkable representation of the experimental
results. This finding of the role of vacancies in the sublattice
responsible for electronic conduction is an important revelation in
the understanding of ferromagnetism in the diluted electron system.}

\begin{document}

\maketitle

\section{Introduction}

Study of ferromagnetism in diluted electron systems (diluted
magnetic semiconductors,\cite{dietl} hexaborides\cite{young} etc.)
having high Curie temperature, $T_C$ has seen an explosive growth in
the recent times due to its potential applications in the spin based
technology. It is believed that in these systems, the spin degrees
of freedom can be used for data storage and data processing in
addition to the use of charge that is used in the present day
technology. Despite numerous studies, the primary concern of the
origin of ferromagnetism in these systems is still unresolved. While
this is a fundamental issue, it is expected to be instrumental for
tailoring new materials those are necessary for the further
development in technology.

In this regard, the discovery of unusual ferromagnetism in
hexaborides, $M$B$_6$ ($M$ = Ca, Sr, Ba etc.) and in La-doped
CaB$_6$ (T$_C \geq$ 600~K) is unique and unusual. High purity
CaB$_6$ is a band insulator.\cite{young,vonlanthen,ott} A small
amount of impurity leads to ferromagnetism in CaB$_6$. While this
fact is somewhat similar to the diluted magnetic semiconductors, the
unique feature in hexaborides is that the magnetization appears
despite the absence of any magnetic element in the composition of
these compounds. Thus, the origin of magnetic moment and hence,
ferromagnetism is puzzling.

Several studies have been performed attributing the ferromagnetism
to the polarization of low density conduction
electrons,\cite{young,ceperley} excitons formed by the doped
holes,\cite{zhitomirsky} magnetic impurity due to the crucibles used
for sample preparation,\cite{matsubayashi} impurities in the boron
sublattice\cite{rhyee} {\it etc}. A recent photoemission
study\cite{cab6PRL} reveals the presence of weakly localized states
in the vicinity of the Fermi level, $\epsilon_F$ in the
ferromagnetic CaB$_6$ while it is absent in the paramagnetic
LaB$_6$.\cite{lab6APL} These studies\cite{rhyee,cab6PRL} show that
ferromagnetism in hexaborides may not be extrinsic arising due to
the magnetic impurities as signature of such elements was not found
in the ferromagnetic compositions. Thus, unlike diluted magnetic
semiconductors, the weak local moment in this system may be related
to the introduction of partial local character in the impurity
states in the vicinity of Fermi level. However, the origin and
character of such localized states are far from understood.

In this letter, we report our results on the influence of impurities
in the electronic structure of CaB$_6$ using state-of-the-art full
potential {\em ab initio} band structure calculations. We observe
that the impurity contributions due to oxygen and carbon at the
boron sites do not play significant role in the magnetic moment
formation. On the other hand, the boron vacancy leads to the
formation of distinct impurity feature in the vicinity of the Fermi
level. The magnetic moment of these features in the spin polarized
calculations are remarkably similar to that found experimentally.

\section{Calculational details}

CaB$_6$ forms in cubic structure with the space group $Pm3m$ and
lattice constant $a$ = 4.152 \AA.\cite{JSSC} In order to introduce
impurities in the B-sublattice, we have doubled the unit cell along
one axis. The unit cell used for calculation is shown in Fig.~1. We
have replaced B12 shown in the figure by C or O impurities
separately, which introduces one impurity among 12 borons (8.3\%
impurity in one formula unit). Vacancies in the B-sublattice are
introduced by removing B12. This allows to introduce 8.3\% defects
in the B-sublattice. Due to the limitation in computational
facilities, larger unit cell could not be considered, which could
allow even lower substitutional levels. Although the experimental
cases correspond to somewhat lower degree of impurities, we believe
that this model provides a good representation of the experimental
cases to find the trend.

Band structure calculations were carried out using full potential
linearized augmented plane wave method within the local spin density
approximations (LSDA) using {\scriptsize WIEN2k}
software.\cite{wien} The energy convergence was fixed to $\Delta E
<$ 0.7~meV/fu (fu = formula unit) and charge convergence achieved
was $\Delta Q <$ 10$^{-3}$ electronic charge/fu. In order to achieve
the convergence 1000 $k$ points were considered within the first
Brillouin zone.

\section{Results and Discussions}

The calculated density of states (DOS) corresponding to CaB$_6$,
CaB$_{5.5}$O$_{0.5}$ and CaB$_{5.5}$C$_{0.5}$ are shown in Fig.~2.
The total density of states (TDOS) in CaB$_6$ exhibit a gap of about
0.2 eV at the Fermi level, $\epsilon_F$ (denoted by zero in the
energy scale) indicating a band insulating phase in the ground
state. The energy range -9 eV to -6 eV is primarily contributed by B
2$s$ states (not shown here to maintain clarity in the figure) and
the energy bands between -6 eV to $\epsilon_F$ have dominant B 2$p$
character (see Fig. 2(c)). Ca 3$d$ bands appear above $\epsilon_F$
with negligible contribution in the occupied part. A rescaling of
the Ca 3$d$ partial density of states (PDOS) by 30 times (shown by
shifting along the y-axis in Fig. 2(b)) reveals that the energy
distribution of Ca 3$d$ states is strikingly similar to that
observed for B 2$p$ states in Fig. 2(c). This indicates significant
covalency between Ca 3$d$ and B 2$p$ states.

The C or O substitution leads to a large shift (by more than 2 eV)
of the 2$p$ bands towards higher binding energies. This is expected
as the total electron count in C and O is more than that in B.
Hence, the Fermi level is expected to shift towards higher energies.
Subsequently, the energy range near the lower energy tail of the Ca
3$d$ bands becomes more intense and partially populated leading the
ground state to a metallic phase. The DOS near $\epsilon_F$ is weak
and flat in the C substituted compound. O substitution generates
relatively larger contribution in the vicinity of $\epsilon_F$ along
with a dip at $\epsilon_F$ and a small but sharp feature at about
-0.5 eV.

The shift of the Ca 3$d$ PDOS appear to be much smaller than that
observed for B 2$s$ and 2$p$ PDOS. The rescaled Ca 3$d$ PDOS in the
lower energy region exhibit DOS distribution similar to that
observed for B 2$s$2$p$ states. The separation between the B 2$p$
PDOS peak and Ca 3$d$ PDOS peak enhances by about 1 eV in the
substituted compounds indicating signature of stronger
hybridization. C 2$p$ and O 2$p$ PDOS looks significantly different
from the distribution observed in B 2$p$ and Ca 3$d$ cases.

In order to investigate the influence of these features in the
ferromagnetic ground state, we have calculated the ground state
energies and magnetic moments for the same lattice within LSDA. We
find that the difference between the total energies corresponding to
non-magnetic and ferromagnetic solutions is less than the error bar
considered for convergence criteria. The magnetic moment at each B
and Ca site is $\sim$~10$^{-3}$~$\mu_B$ per atom or lower. The
magnetic moment centered at the C and O sites are 0.0036~$\mu_B$ per
substitution and 0.007~$\mu_B$ per substitution, respectively. The
total magnetic moment is found to be $<$~0.005~$\mu_B/fu$. The
calculations for CaB$_6$ also exhibit similar scenario. All these
results suggest that the C or O impurities at the boron sites do not
have significant influence in deriving the magnetic moment in this
system. The features in the vicinity of $\epsilon_F$ observed in all
these cases are very different from those observed
spectroscopically.\cite{cab6PRL}

It is thus evident that the magnetic moment observed in CaB$_6$,
although weak, has different origin. We now turn to case of
influence of vacancy in the electronic structure. Removal of B12
from the unit cell introduces 8.3\% vacancy in one formula unit. The
calculated DOS for this case is compared with the results of CaB$_6$
in Fig. 3. The DOS in the energy range away from $\epsilon_F$ (below
-2 eV energy) remains almost similar to the results of CaB$_6$.
Interestingly, the DOS in the vicinity of $\epsilon_F$ exhibit
remarkable spectral weight redistribution. The DOS between 0 to -2
eV energies vanishes and a new energy band appears at $\epsilon_F$
(between -1 to +1 eV). We term this new energy band formed due to
the boron vacancy in the boron sublattice as an {\it impurity band}.
The electronic states corresponding to this {\it impurity band} are
sufficiently itinerant having a bandwidth close to 2 eV and is
partially filled, which indicates that the ground state is metallic.

Although the dominant contributions of the Ca 3$d$ band in
CaB$_{5.5}$ appears in the same energy range as that of CaB$_6$, the
low energy contributions in the vicinity of $\epsilon_F$ exhibit
redistributions similar to the B 2$p$ states. The dominant
contribution in the impurity band comes from the 2$p$ electronic
states corresponding to borons around the vacancy site (B7 - B11) as
shown in Fig. 3(d). The occupied part is primarily contributed by
(B8 - B11) and contributions from B7 appears largely above
$\epsilon_F$.

The total energy for the ferromagnetic solution in this case also
found to be close to that for the non-magnetic solutions; the energy
difference is smaller than the convergence limit as observed in
other cases. Interestingly, the magnetic moment in this case is
found to be significantly larger than all the previous cases. The
magnetic moment centered at B7 site and other four boron sites (B8 -
B11) are 0.023~$\mu_B$ and 0.003~$\mu_B$ respectively. The magnetic
moment at all the other sites are very small. The total magnetic
moment is found to be 0.018~$\mu_B/fu$, which is almost identical to
that found experimentally.\cite{lofland}

In Fig. 4, we show the calculated spin polarized DOS. It is evident
that the exchange splitting is not distinctly visible in TDOS and Ca
3$d$ partial DOS (PDOS). However, the 2$p$ contributions from B7 and
(B8 - B11) in the {\it impurity band} exhibit signature of finite
exchange splitting. The splitting is maximum ($\sim$~0.25 eV) in the
case of B7 (see Fig. 4(b)). This is also manifested in the magnetic
moment. This indicates that the 2$p$ electronic states corresponding
to B7 site have significant local character and play a key role in
the ferromagnetism of these materials.

All the above results are remarkably consistent with the
experimental observation of the 2$p$ character of the impurity
feature in the high resolution photoemission study\cite{cab6PRL} and
the observation of the strong dependence of magnetic moment on
defects.\cite{lofland} The magnetism involving 2$p$ electrons has
been predicted in other systems such as C impurities in BN
nanotubes.\cite{Wu} Oxygen vacancy induced ferromagnetism has also
been observed in $d^0$ oxides.\cite{bouzerar} In this study, we
observe that the C and O impurities do not play significant role in
magnetic moment formation in CaB$_6$. The defects states due to the
B vacancy form an {\it impurity band} in these systems. The {\it
impurity band} has significant bandwidth to ensure itineracy of the
associated electrons and local character that plays the dominant
role in determining the magnetic moment. These results, thus,
provide an important input in the understanding of magnetism of
these interesting materials. We hope that this will help to initiate
further studies in this direction that is necessary to design and
fabricate new materials of potential technological importance.

\section{Conclusions}

In summary, we have investigated the origin of magnetic moment
formation in CaB$_6$ using state-of-the-art full potential band
structure calculations. CaB$_6$ is found to be a band insulator with
a band gap of about 0.2 eV. Introduction of carbon or oxygen leads
to a shift of the Fermi level towards higher energies. In these
cases the ground state is metallic. The Ca 3$d$ and B 2$p$
hybridization strength is found to increase due to such
substitutions. No influence of C or O substitution was observed in
the formation of magnetic moment.

Introduction of vacancy in the boron sublattice leads to the
formation of an {\it impurity band} in the vicinity of the Fermi
level. Such impurity states have B 2$p$ character consistent with
the experimental observations and are located close to the vacancy
site. The exchange splitting of these states is found to be finite
leading to a magnetic moment that is consistent with the
experimental results. These results suggest that the boron vacancy
plays the key role in ferromagnetism in these diluted electron
systems.

%
%
%

%

\begin{figure}
\vspace{-4ex}
 \centerline{\epsfysize=7.0in \epsffile{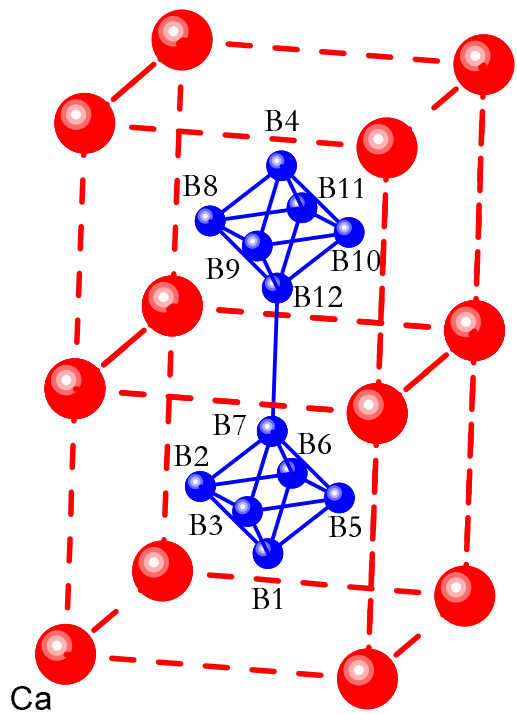}}
\vspace{-2ex}

 \caption{(color online) The unit cell considered for the calculations
containing 2 formula unit of CaB$_6$. All the borons are labeled to
identify the borons around the defect site. B12 is removed/replaced
to introduce vacancy/impurity.}
\end{figure}

\begin{figure}
\vspace{-4ex}
 \centerline{\epsfysize=7.0in \epsffile{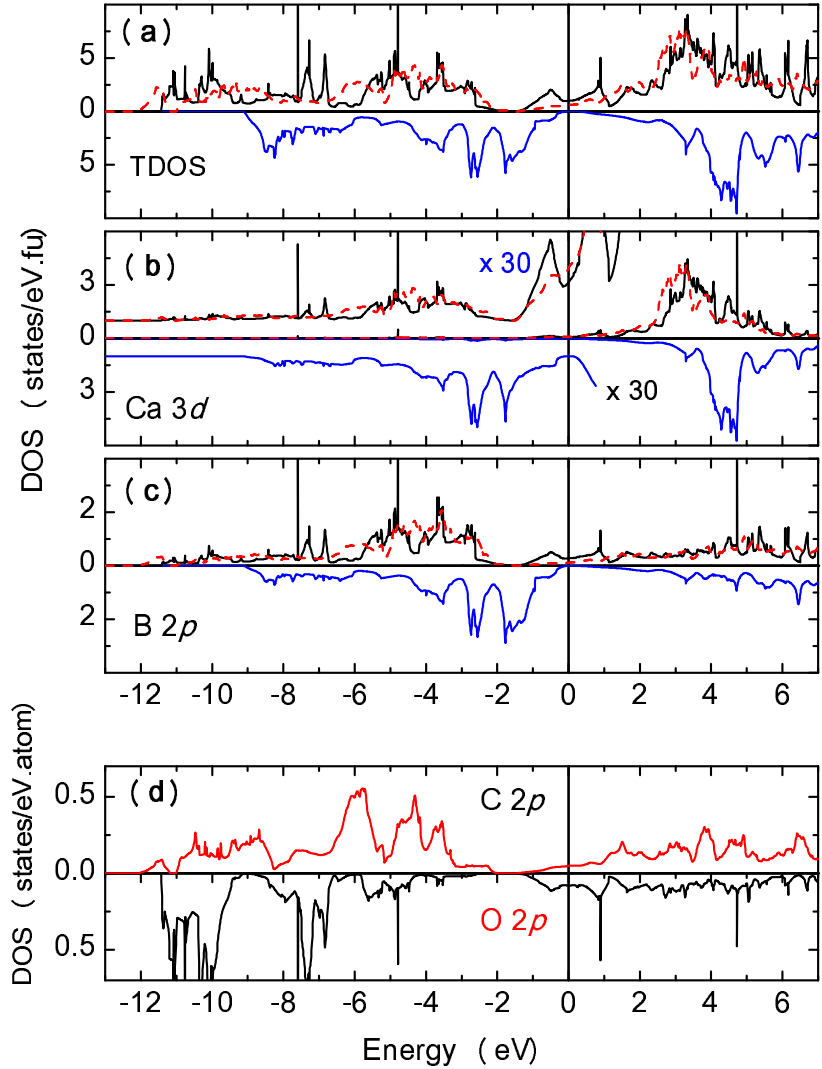}}
\vspace{-2ex}
  \caption{(color online) Calculated (a) total density of states (TDOS)
(b) Ca 3$d$ partial density of states (PDOS), and (c) B 2$p$ PDOS
are shown. The results corresponding to CaB$_6$ are shown by
inverted $y$-axis. Along the positive $y$-axis, CaB$_{5.5}$C$_{0.5}$
(dashed line) and CaB$_{5.5}$O$_{0.5}$ (solid line) are shown. In
order to visualize Ca 3$d$ PDOS in low energy region, (Ca 3$d$
PDOS~$\times$~30~+~1) are shown in (b). (d) C 2$p$ PDOS (positive
$y$-axis) and O 2$p$ PDOS (inverted axis) for CaB$_{5.5}$C$_{0.5}$
and CaB$_{5.5}$O$_{0.5}$, respectively.}
\end{figure}

\begin{figure}
\vspace{-4ex}
 \centerline{\epsfysize=7.0in \epsffile{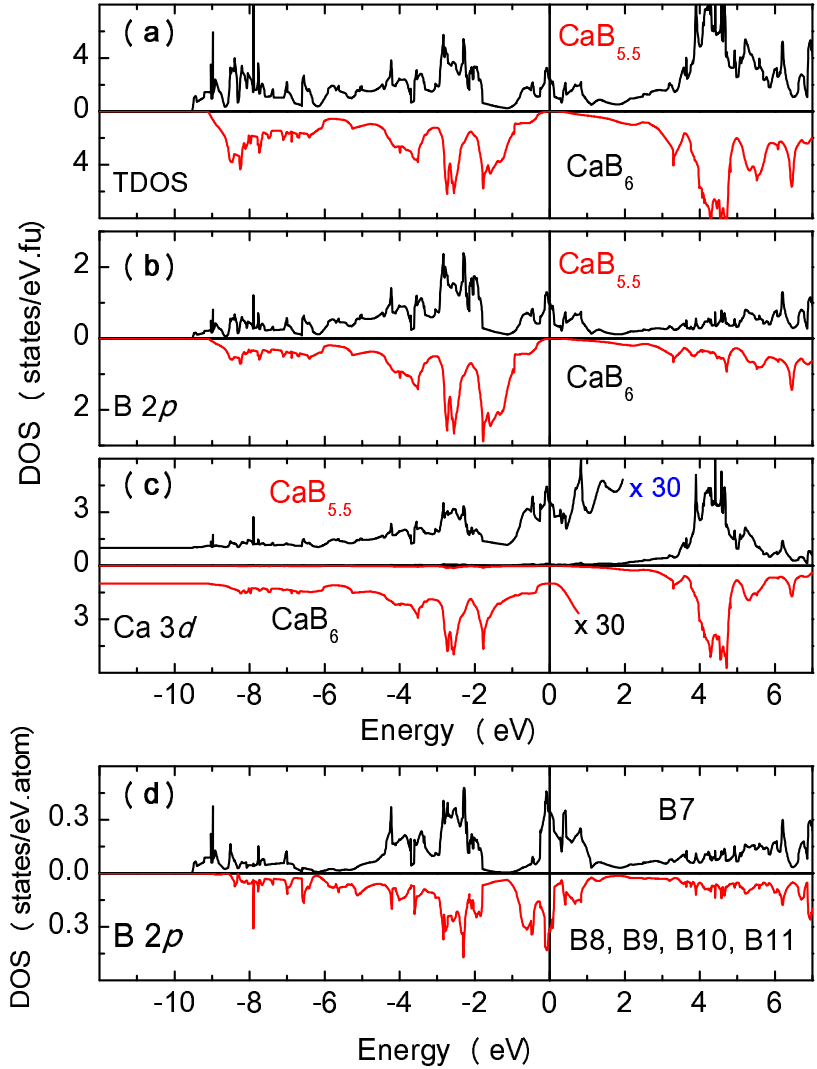}}
\vspace{-2ex}
 \caption{(color online) Calculated (a) total density of states
(TDOS), (b) B 2$p$  partial density of states (PDOS), and (c) Ca
3$d$ PDOS are shown. The results corresponding to CaB$_6$ are shown
by inverted $y$-axis and those of CaB$_{5.5}$ are shown along the
positive $y$-axis. In order to visualize Ca 3$d$ PDOS in low energy
region, Ca 3$d$ PDOS~$\times$~30~+~1 are shown in (c). (d) 2$p$ PDOS
corresponding to B7 and B8 in CaB$_{5.5}$ are shown by positive
$y$-axis and inverted $y$-axis, respectively. 2$p$ PDOS
corresponding to B9, B10 and B11 are identical to that of B8.}
 \label{fig.1}
\end{figure}

\begin{figure}
\vspace{-4ex}
 \centerline{\epsfysize=7.0in \epsffile{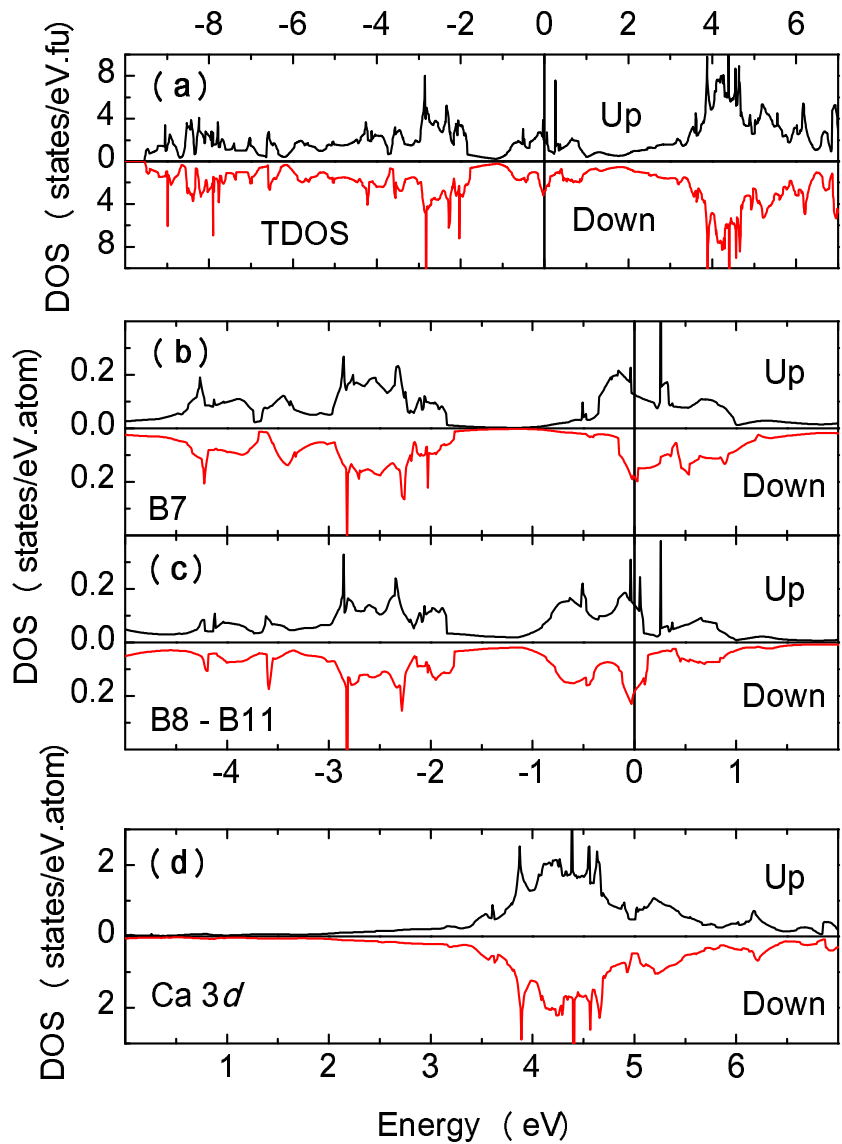}}
\vspace{-2ex}
 \caption{(color online) Spin polarized density of states
corresponding to (a) TDOS, (b) B7 2$p$ PDOS, (c) 2$p$ PDOS of
(B8-B11), and (d) Ca 3$d$ PDOS are shown with up-spin component
along positive $y$-axis and down-spin component along inverted
$y$-axis. Clearly, 2$p$ PDOS corresponding to B7 has the largest
exchange splitting and hence large magnetic moment compared to all
other borons.}
\end{figure}

\end{document}